\documentclass[twocolumn,prl,aps,superscriptaddress]{revtex4}
\usepackage[T1]{fontenc}
\usepackage{graphicx}
\newcommand{\beq}[2]{\begin{equation}#1\label{#2}\end{equation}}
\newcommand{\ceq}[1]{(\ref{#1})}
\newcommand{\mbd}[1]{\mbox{\bf #1}}

\newcommand{\br}{\mbd{r}}
\newcommand{\bx}{\mbd{x}}
\newcommand{\by}{\mbd{y}}

\newcommand{\bJ}{\mbd{J}}
\newcommand{\bA}{\mbd{A}}
\newcommand{\bB}{\mbd{B}}
\newcommand{\bC}{\mbd{C}}
\newcommand{\bD}{\mbd{D}}

\newcommand{\bxi}{\BF{\xi}}

\newfont{\mbld}{cmbx10 scaled 800}
\newfont{\cab}{cmsy10 scaled 1200}
\newfont{\scab}{cmsy10 scaled 1000}
\newfont{\bcall}{cmbsy10 scaled 1200}

\newcommand{\BF}[1]{\mbox{\boldmath $#1$}}          
\newcommand{\nablab}{\BF{\nabla}}
\begin{document}
\title{Topologically Linked Polymers are Anyon Systems}
\author{Franco Ferrari}
\email{ferrari@univ.szczecin.pl}
\affiliation{Institute of Physics, University of Szczecin,
  ul. Wielkopolska 15, 70-451 Szczecin, Poland}
\affiliation{Max Planck Institute for Polymer Research, 10
  Ackermannweg, 55128 Mainz, Germany}

\begin{abstract}
We consider the
statistical mechanics of a system of topologically linked polymers,
such as for
instance a dense solution of polymer rings.
If the possible topological states of the system are distinguished
using the Gauss linking number as a topological invariant, 
the partition function of an ensemble of
$N$ closed polymers coincides with the $2N$ point function of a field theory
containing a set of $N$ complex replica fields
and Abelian Chern-Simons  fields. Thanks to this
mapping
to field theories,
some quantitative predictions
on the behavior of topologically entangled polymers have been
obtained by exploiting perturbative techniques. In order to go
beyond perturbation theory, a connection
between polymers and anyons  is established here.
It is shown 
in this way that the topological forces which
maintain two polymers in a given topological configuration have
both attractive and repulsive components.
When these opposite components reach
a sort of equilibrium, the system finds itself in a
self-dual point
similar to that which, in the Landau-Ginzburg model for
superconductors, corresponds to the transition from type I to type II
superconductivity.
The significance of self-duality in polymer physics is
illustrated considering the example of the so-called
$4-plat$ configurations, which
are of interest in the biochemistry of DNA processes like replication,
transcription and recombination. The case of static
vortex solutions of the Euler-Lagrange equations is discussed.
\end{abstract}
\maketitle
In this letter we consider the
statistical mechanics of a system of topologically linked
polymers\cite{khovil},
such as for
instance a dense solution of polymer rings (catenanes).
If the possible topological states of the system are distinguished
using the Gauss linking number as a topological invariant, it has been
shown in \cite{FKLnpol} that the partition function of an ensemble of
$N$ closed polymers coincides with the $2N$ point function of a field theory
containing a set of $N$ complex replica fields
and Abelian Chern-Simons (C-S) fields. Thanks to this
mapping
to field theories,
some quantitative predictions
on the behavior of topologically entangled polymers have been
obtained by exploiting perturbative methods, see
Ref.~\cite{Ferrari:ts} and references therein. If one wishes to go
beyond perturbation theory, however, one is faced with the
problem of field theories which contain
 multi-component 
scalar fields and
gauge fields interacting together.
Clearly, it is not easy to study theories of this kind even in the
simplest case $N=2$.
The idea which we wish to pursue here in order to circumvent such
difficulties is to establish a link between polymers and anyons
\cite{Wilczek}.
Anyon field theories have been in fact intensively investigated. It
is known for instance that their classical equations of motion
admit vortex solutions \cite{JackiwPi}.
In a special region of the space of parameters, called
the Bogoml'nyi self-dual point \cite{Bogom},
the attractive and repulsive forces
between vortices 
vanish. This point corresponds in the Landau-Ginzburg model for
superconductors to the transition from type I to type II
superconductivity.

\begin{figure}[bpht]
\centering

\includegraphics[scale=.4]{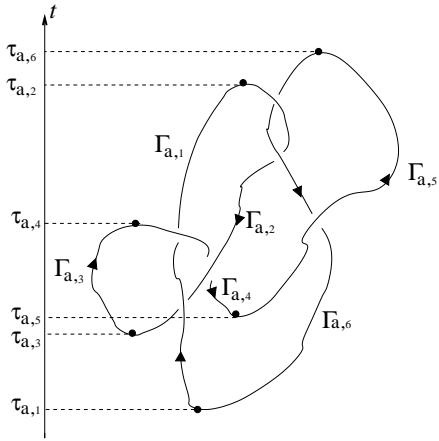}\\
\caption{Sectioning procedure for a $2s-plat$ $\Gamma_a$ with
$s=3$.}\label{sectioning}
\end{figure}

The first obstacle in this program is that anyons are living in $2+1$
dimensions, while polymer trajectories are intrinsecally defined in
three dimensions. For this reason, one coordinate must be singled out
and regarded as {\it time}.
Accordingly, a point of the three-dimensional Euclidean space $\mathbf
R^3$ will be identified by a three vector $\xi^\mu=(\br,t)$, where
$\mu=1,2,3$, $\br=(x^1,x^2)$ and $x_3=t$ plays the role of time.
We need also to introduce the two-dimensional completely antisymmetric
tensor $\epsilon^{ij}$, $i=1,2$ and its three dimensional counterpart
$\epsilon^{\mu\nu\rho}$, $\mu,\nu,\rho=1,2,3$. These tensors are
uniquely defined by the following conventions: $\epsilon^{12}=1$
and $\epsilon^{123}=1$. 
We use middle latin letters as two-dimensional space indices and
middle greek letters as three dimensional indices. Sum over repeated
space indices will be everywhere understood.
The trajectories $\Gamma_a$ of the polymers,
$a=1,\ldots,N$, will be treated as continuous curves. Usually, these
trajectories are parametrized by means of their arc-length $\sigma_a$:
$\Gamma_a=\left\{ (\br_a(\sigma_a),t_a(\sigma_a))\right|\left.
0\le\sigma_a\le L_a\right\}$, $L_a$ being the total length of the
trajectory.
To make a connection with anyons, however, it is convenient
 to use the time $t$ as a parameter. Of
course, the coordinate $t$ is not a good parameter 
if we do not introduce a suitable
of {\it sectioning} for the curves $\Gamma_a$. To this
purpose, let us notice that, with respect to the $t-$direction,
each curve $\Gamma_a$ has maxima
and minima defined by the condition:
$\frac{dt_a(\sigma_a)}{d\sigma_a}=0$. The number of minima, $s_a$,
is equal to the number of maxima. Let us pick up a given point
of minimum on $\Gamma_a$ and call it $\tau_{a,1}$. Starting from
$\tau_{a,1}$ and going along the curve after
choosing arbitrarily a direction,
one will encounter successively a point of
maximum, that will be called $\tau_{a,2}$, a point of minimum
$\tau_{a,3}$ and so on. 
In this way we obtain a set of $2s_a$ points
$\tau_{a,1},\ldots,\tau_{a,2s_a}$.
Now it is possible to split $\Gamma_a$ into $s_a$ open paths
$\Gamma_{a,1},\ldots, \Gamma_{a,s_a}$ which connect pairwise
contiguous points of maxima and minima:
\beq{
\Gamma_{a,I_a}=
\left\{(\br_{a,I_a}(t),t)
{\Bigg |}
\begin{array}{l}
\tau_{a,I}\le t\le\tau_{a,I+1}\\
\br_{a,I_a}(\tau_{a,I_a+1})=\br_{a,I_a+1}(\tau_{a,I_a+1})\\
\br_{a,I_a}(\tau_{a,I_a})=\br_{a,I_a-1}(\tau_{a,I_a})
\end{array} 
  \right\} 
}{gaaI}
where $I_a$ is a cyclic index such that
$I_a\in\left\{1,\ldots,2s_a\right\}$
and $I_a+2s_a=I_a$. The points $\br_{a,I_a}(\tau_{a,I_A})$ 
are considered as fixed points.
This sectioning procedure is explicitly illustrated in
the case $s=3$ by Fig.;~\ref{sectioning}.
By construction, there is a one-to-one correspondence between the
points of the $t-$axis in the interval $[\tau_{a,I},\tau_{a,I+1}]$ and
the points of the trajectories $\Gamma_{a,I}$. As a consequence, it is
possible to parametrize the trajectories $\Gamma_{a,I_a}$ by means of the
variable $t$.

We are now ready to construct the partition function of a system of
topologically entangled polymers. For the sake of clarity, it is
convenient to use some of the nomencleature of knot theory. In
particular, from the mathematical point of view, each closed
trajectory $\Gamma_a$ can be regarded as a knot, while an ensemble of two
or more trajectories defines a link. Moreover, with some slight abuse
of language, a knot or a link with $2s$ points of maxima and minima will be
called a $2s-$plat. 
The concept of plats arises quite naturally in the processes of
replication, recombination and transcription of DNA \cite{wacozsum}.

To start with, let us write the partition function for a link
with the topology of a $2S-$plat, where $S=\sum_{a=1}^n s_a$.
The integers $s_a$ are now kept fixed, so that the link is composed by
knots $\Gamma_a$ which are $2s_a-$plats.
In order to avoid a proliferation of indices, we will restrict
ourselves to a system of two linked polymers, thus setting
$N=2$. Actually, this is not a big limitation, because the
two-polymer problem contains all the ingredients of the $N-$polymer
problem. 
Putting $\imath=\sqrt{-1}$, the desired polymer partition function is
given by:
\beq{
{\cal Z}(\lambda)=\int{\cal D}B_\mu{\cal D}C_\nu
e^{-\imath {\cal S}_{CS}}
\int_{\mbox{\scriptsize boundary}\atop{\mbox{\scriptsize
    conditions}}} {\cal D}\br_{a,I_a}(t)e^{-{\cal A}}
}{ppm}
The boundary conditions on the trajectories $\Gamma_{a,I_a}$ are those
of Eq.~(\ref{gaaI}). These boundary conditions are required by the
fact that the open paths $\Gamma_{a,I_a}$, with
$a$ fixed and $I_a=1,\ldots,2s_a$, must join together in a suitable
way at the points of maxima and minima $\tau_{a,I_a}$ in order to
reconstruct the closed curve $\Gamma_a$.

The polymer action ${\cal A}$ can be divided into two contributions:
\beq{{\cal A}={\cal A}_{free}+{\cal A}_{top}
}{fquantpolact}
The free part ${\cal A}_{free}$ is:
\beq{{\cal A}_{free}=\sum_{a=1}^2\sum_{I_a=1}^{2s_a}
\int_{\tau_{a,I_a}}^{\tau_{a,I_a+1}}
dt(-1)^{I_a-1}g_{aI_a}\left|\frac{d\br_{a,I_a}(t)}{dt}\right|
}{sfree}
The factors $(-1)^{I_a-1}$ appearing in Eq.~(\ref{sfree})
are necessary to make ${\cal A}_{free}$ positive definite. The
parameters $g_{aI_a}$ are related to the lengths $L_{a,I_a}$ of the open chains
$\Gamma_{a,I_a}$ as follows. Let us discretize
$\Gamma_{a,I_a}$ considering it as a random chains with $n$ segments.
If Eq.~(\ref{sfree}) would be the action of a set of free two-dimensional
chains,
in the limit of large values of $n$ and keeping fixed the
ratio $n/g_{a,I_a}$, one would obtain the well known result
\cite{Kleinertbook}:
$L_{a,I_a}\sim n/ g_{a,I_a}$. Here things are a little bit more
complicated, because one has to take into account the third
dimension, which in turn is also the variable which parametrizes the curves.
As a consequence, the above relation governing the length of the
trajectories is replaced by:
\beq{L_{a,I_a}^2\sim |\tau_{a,I_a+1}-\tau_{a,I_a}|^2+
\frac n{g_{a,I_a}}|\tau_{a,I_a+1}-\tau_{a,I_a}|
}{lengthbasicrel}
The second term in Eq.~(\ref{fquantpolact}) is:
\[
{\cal A}_{top}=\imath \lambda
\sum_{I=1}^{2s_1}\int_{\tau_{1,I}}^{\tau_{1,I+1}} dt
\frac{d\xi^\mu_{1,I}(t)}{dt} B_\mu(\br_{1,I}(t),t)
\]
\beq{+\imath \frac \kappa{2\pi} \sum_{J=1}^{2s_2}dt
\frac{d\xi^\mu_{2,J}(t)}{dt} C_\mu(\br_{2,J}(t),t)
}{stop}
In the above equation $B_\mu=(\bB,B_0)$ and $C_\mu=(\bC,C_0)$ are two
Abelian C-S vector fields. We have put $B_3\equiv B_0$ and
$C_3\equiv C_0$ in order to stress the role of time of the third spatial
coordinate. $\kappa$ and $\lambda$ are real parameters. $\kappa$
coincides with the C-S coupling constant (see below).
The spurious gauge degrees of freedom have been eliminated using the
gauge condition (Coulomb gauge):
\beq{\nablab\cdot\bB=\nablab\cdot\bC=0}{coulgauge}
so that the gauge fixed C-S action $S_{CS}$ looks as follows:
\beq{S_{CS}=\frac\kappa{4\pi} \int d^3\xi\left[
B_0\epsilon^{ij}\partial_iC_j+C_0\epsilon^{ij}\partial_iB_j
\right]}{csacg}
The C-S fields act as auxiliary fields which impose topological
constraints on the closed trajectories $\Gamma_1$ and $\Gamma_2$.
To show this, one has to integrate out these fields from the
partition function (\ref{ppm}) and then to perform an inverse Fourier
transformation of ${\cal Z}(\lambda)$:
\beq{{\cal Z}(m)=\int_0^{2\pi}\frac{d\lambda}{2\pi}e^{im\lambda}{\cal
    Z}(\lambda) }{invFT}
The details of this calculation
 have been already explained in
Ref.~\cite{Ferrari:ts} and will not be repeated here.
The result is the new partition function ${\cal Z}(m)$, which
describes the fluctuations of two polymer rings constrained by the
condition:
\beq{m=\chi(\Gamma_1,\Gamma_2)}{topconstr}
where $\chi(\Gamma_1,\Gamma_2)$ is the Gauss linking number and the
topological number $m$ must be an integer.
A few comments are in order at this point. First of all,
in \cite{Ferrari:ts} 
the topological constraint (\ref{topconstr}) has been derived from
the C-S path integral using the Lorentz gauge, while here 
the Coulomb gauge (\ref{coulgauge}) has been chosen. Formally, things
works exactly in the same way in both gauges. However,
one should keep in mind that the final
expression of the Gauss linking number $\chi(\Gamma_1,\Gamma_2)$
obtained starting from the Lorentz gauge differs from the expression
that one obtains starting
from the
Coulomb gauge. Of course, the two expressions are equivalent due to
gauge invariance, but the way in which the winding of one trajectory
around the other is counted, is realized in a very different way.
 Another important point is that
our
approach allows a more refined classification of links than the bare
Gauss linking number. Indeed, two links 
which have the same Gauss linking number can be further distinguished
by their plat-structure, i.~e. by the number of maxima and minima
of the knots which are composing the links.
Finally, it is not difficult to add
also the repulsive steric interactions between the
monomers in the partition function
(\ref{ppm}). It is interesting to notice that,
since we have chosen
the coordinate $x_3\equiv t$ in order to parametrize the curves,
the usual $\delta-$function potential which describes
these forces takes the form:
$V(\br,t)=v_0\delta(\br)\delta(t)$, $v_0$ being a temperature
dependent constant.
Due to the presence of the $\delta-$function in time,
the steric interactions become instantaneous in our case.
In the following we will not discuss
 the steric interactions, in part because we prefer to concentrate
on the pure topological interactions, in part because the inclusion of
these repulsive forces
complicates enormously the task of
finding non-perturbative analytic solutions of the two-polymer problem under
consideration.
Thus, we will assume that in our system steric interactions are
negligible. This approximation is valid for instance in polymer
solutions which are very dilute or in which the temperature is near
the so-called theta-point. The hypothesis of very low monomer densities
is natural in the present context, because we are discussing
the statistical properties of two isolated polymer rings, which are
supposed to be very long and thin.

To establish the connection with anyons, we need to pass in the
partition function (\ref{ppm}) from the statistical sum over the
trajectories $\br_{a,I_a}(t)$ to a path integral over fields. 
 The details of the derivation of the partition function ${\cal
  Z}(\lambda)$ in terms of fields in the general case and including
also the steric interactions will be presented elsewhere. Here we will
restrict ourselves to links which have the structure of $4-$plats,
putting henceforth $s_1=s_2=1$. This case is particularly interesting for
biological applications, since most of the links produced by in vitro
enzymology experiments are $4-$plats \cite{wacozsum}.
Moreover, ``small'' links up to
seven crossings are all $4-plats$.

Introducing a set of
$n_r$ complex replica fields:
\beq{\begin{array}{rcl}
{\vec \Psi}_{a,I_a}&=&(\psi_{a,I_a}^1,\ldots,\psi_{a,I_a}^{n_r})\\
{\vec \Psi}_{a,I_a}^*&=&(\psi_{a,I_a}^{1*},\ldots,\psi_{a,I_a}^{n_r*})
\end{array}
}{replfields}
and a convenient notation for covariant derivatives:
\beq{
\bD(\pm\lambda,\bB)=\nablab\pm\imath\lambda\bB\qquad
\bD(\pm\frac{\kappa}{2\pi},\bC)=\nablab\pm\imath\frac{\kappa}{2\pi} \bC
}{covdevsdef}
the $4-$plat partition function can be expressed as follows:
\beq{
{\cal Z}_{4-plat}(\lambda)=\lim_{n\to 0}\int{\cal D}(fields)\prod_{I=1}^{2}
{\cal O}_{1,I}
\prod_{J=1}^{2}{\cal O}_{2,J}e^{-{\cal S}_{4-plat}}
}{anyonsppm}
The action ${\cal S}_{4-plat}$ contains the free field action and the
topological terms:
\begin{widetext}
\begin{eqnarray}
{\cal S}_{4-plat}
&=&
\int\limits_{\tau_{1,1}}^{\tau_{1,2}}dt\int
d\bx\left\{
{\vec \Psi}_{1,1}^*{
\partial_t}{\vec
  \Psi}_{1,1}
+{ \frac1{4g_{11}}}\left|\bD(-\lambda,\bB){\vec \Psi}_{1,1}
\right|^2+
{\vec \Psi}_{1,2}^*{\partial_t}{\vec
  \Psi}_{1,2}
+{ \frac1{4g_{12}}}\left|\bD(\lambda,\bB){\vec \Psi}_{1,2}
\right|^2\right\}+\nonumber\\
&+&
\int\limits_{\tau_{2,1}}^{\tau_{2,2}}dt\int
d\bx\left\{
{\vec \Psi}_{2,1}^*{\partial_t}{\vec
  \Psi}_{2,1}+
{ \frac1{4g_{21}}}\left|\bD(-
{\frac{\kappa}{2\pi}},\bC){\vec \Psi}_{2,1}
\right|^2+
{\vec \Psi}_{2,2}^*{\partial_t}{\vec
  \Psi}_{2,2}
+{ \frac1{4g_{22}}}\left|\bD({\frac{\kappa}{2\pi}}
,\bC){\vec \Psi}_{2,2}
\right|^2\right\}\label{anyonsaction}
\end{eqnarray}
\end{widetext}
$d\bx\equiv dx_1dx_2$ is the infinitesimal volume element in the
two-dimensional space and $\partial_t=\frac\partial{\partial t}$.
Let us note that in the partition function (\ref{anyonsppm}) we have
already integrated out the time components of the Chern-Simons fields
$B_0,C_0$, which play the role of Lagrange multipliers and impose the
following constraints:
\begin{eqnarray}
{\cal
  B}=2\left(\left|{\vec\Psi}_{2,1}\right|^2-\left|{\vec\Psi}_{2,2}\right|^2
  \right) &\theta(\tau_{2,2}-t)\theta(t-\tau_{2,1})&\label{constrone}\\
{\cal
  C}={ \frac{4\pi}{\kappa}}
\left(\left|{\vec\Psi}_{1,1}\right|^2-\left|{\vec\Psi}_{1,2}\right|^2
  \right) &\theta(\tau_{1,2}-t)\theta(t-\tau_{1,1})&\label{constrtwo}
\end{eqnarray}
In the above equation
 ${\cal B}$ and  ${\cal C}$ are the magnetic fields associated to
the vector potentials $\bB$ and $\bC$ respectively:
\begin{eqnarray}
{\cal
  B}&=&\partial_1B_2-\partial_2B_1=\epsilon^{ij}\partial_iB_j\label{magb}\\
{\cal C}&=&\partial_1C_2-\partial_2C_1=\epsilon^{ij}\partial_iC_j\label{magc}
\end{eqnarray}
and $\theta(t)$ is the Heaviside theta function $\theta(t)=0$ if $t<0$
and $\theta(t)=1$ if $t\ge 0$.
The symbol ${\cal D}(fields)$ in Eq.~(\ref{anyonsppm}) is a shorthand
for the field integration measure:
\beq{
{\cal D}
(fields)=\int\prod_{I_a=1}^{2s_a}{\cal
  D}{\vec\Psi}_{a,I_a}^*
{\cal D}{\vec \Psi}_{a,I_a}
}{measfield}
Finally, the operators ${\cal O}_{a,I_a}$
are given by:
\begin{eqnarray}
{\cal O}_{a,I_a}&=&
\psi_{a,I_a}^1(\br_{a,I_a}(\tau_{a,I_a+1}),
\tau_{a,I_a+1})\nonumber\\
&&\psi_{a,I_a}^{1*}(\br_{a,I_a}(\tau_{a,I_a}),
\tau_{a,I_a})\label{operators}
\end{eqnarray}
Eqs.~(\ref{anyonsppm})--(\ref{operators}) establish the desired
connection between polymers and anyons. In fact, the action
(\ref{anyonsaction}), together with the constraints
(\ref{constrone})--(\ref{constrtwo}), coincides formally with the
action of a multi-layered anyon system. The number of layers is
$2(s_1+s_2)=4$, while the fields ${\vec \Psi}_{a,I_a}^*,{\vec
  \Psi}_{a,I_a}$  propagate anyonic particles with spin $n_r$ and the
statistics of boson. One difference from anyonic field
theories is due to the fact that, in the case of real particles, one
considers the evolution of the whole system from an initial time $T_1$
and a final time $T_2$. Here, instead, time intervals measure
 the extensions of
the trajectories $\Gamma_{a,I_a}$ in the $x_3$ direction. For this
reason, in the partition function (\ref{anyonsppm}) part of the system
evolves within the time interval $[\tau_{1,1},\tau_{1,2}]$ and another
part within the time interval $[\tau_{2,1},\tau_{2,2}]$. This
inhomogeneity in the time intervals and the use of the Coulomb gauge,
which makes the interactions instantaneous, is the cause of the
presence of the Heaviside $\theta-$functions in
Eqs.~(\ref{constrone})--(\ref{constrtwo}). In this sense, the complete
analogy with anyons can be recovered only in the limit
\beq{\tau_{1,1}=\tau_{2,1}\equiv T_1\qquad
  \tau_{1,2}=\tau_{2,2}=T_2}{equalmaxmin} 
Throughout the rest of this letter we will assume that the above
condition is satisfied. This is to avoid technical complications in
dealing with the $\theta-$functions.

The connection to the anyon problem
suggest to apply
to the action of Eq.~(\ref{anyonsaction}) 
the Bogomol'nyi transformations \cite{Bogom}.
Using these transformation in a suitable way, 
it is
possible to rewrite  ${\cal S}_{4-plat}$
in the following form:
\beq{{\cal S}_{4-plat}={\cal F}_{4-plat}+{\cal I}_{4_plat}}{sdpnsd}
Neglecting surface
terms and introducing the notation 
$D_\pm=D_1\pm \imath D_2$, where $D_i$, $i=1,2$,
denotes the $i-$th space components of the covariant derivatives
(\ref{covdevsdef}), one has that:
\begin{widetext}
\begin{eqnarray}
&&{\cal F}_{4-plat}
=\int_{T_1}^{T_2}
dt\int d\bx\left[
\sum_{a=1}^2\sum_{I_a=1}^2{\vec \Psi}_{a,I_a}^*\partial_t{\vec
    \Psi}_{a,I_a}+\right.\nonumber\\
&&+
{ \frac1{4g_{11}}}\left|D_+(-\lambda,\bB){\vec \Psi}_{1,1}\right|^2+
{
\frac1{4g_{12}}}\left|D_+(\lambda,\bB){\vec \Psi}_{1,2}\right|^2+
\left.
{
\frac1{4g_{21}}}\left|D_-(-{ \frac{\kappa}{2\pi}}
,\bC){\vec \Psi}_{2,1}\right|^2+
{ \frac1{4g_{22}}}\left|D_-({\frac{\kappa}{2\pi}}
,\bC){\vec \Psi}_{2,2}\right|^2\right]\label{sdaction}
\end{eqnarray}
and
\begin{eqnarray}
&&{\cal I}_{4-plat}=\frac\lambda 2
\int_{T_1}^{T_2}dt
\int d\bx
\left[
\left({\textstyle
\frac{
\left|
{\vec\Psi}_{1,1}
\right|^2}
{g_{11}}}-
{\textstyle \frac{
\left|
{\vec\Psi}_{1,2}
\right|^2}
{g_{12}}}\right)
\times
{\textstyle \left(
\left|
{\vec\Psi}_{2,1}
\right|^2-
\left|
{\vec\Psi}_{2,2}
\right|^2
\right)}
-\left({\textstyle\frac
{\left|{\vec\Psi}_{2,1}\right|^2}{g_{12}}}
-{\textstyle\frac
{\left|{\vec\Psi}_{2,2}\right|^2}{g_{22}}}
\right)\times
\left(
\left|{\vec\Psi}_{1,1}\right|^2
-
\left|{\vec\Psi}_{1,2}\right|^2
\right)
\right]\nonumber\\
\label{coulpots}
\end{eqnarray}
\end{widetext}
It is easy to recognize from Eq.~(\ref{sdaction})
that ${\cal
  F}_{4-plat}$ consists in the sum of the self-dual actions of four
different families of anyons ${\vec \Psi}_{a,I_a}$.
These families are coupled together by the
constraints (\ref{constrone})--(\ref{constrtwo}).
The second contribution to ${\cal S}_{4,plat}$, the interaction term
${\cal I}_{4-plat}$, contains
instead Coulomb-like potentials, which attract or repel the
monomers belonging to different trajectories $\Gamma_{a,I_a}$.
It is difficult to guess what is the influence of these Coulomb
forces
on the statistical behavior of the polymers. The problem is
that their strength is proportional to $\lambda$ (see
Eq.~(\ref{coulpots})) and $\lambda$ is not a real coupling constant
with a physical meaning. At
the end, we are interested to eliminate $\lambda$ and to
express the polymer partition function in terms of the topological
number $m$, as we did in Eq.~(\ref{invFT}). $m$ is in fact the true
physical parameter which describes the topological state of the
system.
What is possible to conclude from Eqs.~(\ref{sdpnsd})--(\ref{coulpots}) is
that topological forces have attractive and repulsive components, which
interfere with the steric interactions. This result confirms at a
non-perturbative level a previous result obtained using perturbative
methods \cite{Ferrari:ts}. 
Also in Ref.~\cite{witten} it has been shown
in the limit in which fluctuations in the particle densities are relatively
small that
the C-S fields
generate effective Coulomb interactions 
in anyon models.

Remarkably, our approach reveals that the
two-polymer system under consideration has a self-dual point. This
point is reached when the parameters $g_{aI_a}$ are equal:
\beq{g_{11}=g_{12}=g_{21}=g_{22}=g}{equalgaia}
If the above relations are satisfied, it is easy to see that
${\cal I}_{4-plat}=0$, so that the only surviving terms in the action
${\cal S}_{4-plat}$ are those coming from the self-dual contributions
in ${\cal F}_{4-plat}$.
We note that
the existence of the self-dual
point \ceq{equalgaia} is not submitted to the assumption
of Eq.~\ceq{equalmaxmin}.
Indeed,  let us suppose that the points of maxima and minima
$\tau_{a,I_a}$ of the trajectories $\Gamma_{a,I_a}$ are not aligned as
specified by Eq.~\ceq{equalmaxmin}.
The only difference in the expression of the interaction
term ${\cal I}_{4-plat}$ is that now the values of $T_1$ and $T_2$, which
give the boundaries of the integration over time,  should be replaced
by the following ones:
\beq{T_1=\max[\tau_{1,1},\tau_{2,1}]\qquad\qquad
T_2=\min[\tau_{1,2},\tau_{2,2}]}
{gentimeint}
The  choice \ceq{gentimeint}
of integration limits in ${\cal
  I}_{4-plat}$ is dictated by the presence of the $\theta-$functions
of Heaviside in the constraints (\ref{constrone})--(\ref{constrtwo}).
Physically, Eq.~(\ref{gentimeint}) is motivated by the fact that, in
the Coulomb gauge, all interactions become instantaneous. Therefore, if
maxima and minima are not aligned,  it is easy to see that the
trajectories may only interact in the time interval $[T_1,T_2]$, where
$T_1$ and $T_2$ are given by Eq.~(\ref{gentimeint}).
Substituting the new integration limits \ceq{gentimeint}
 in
Eq.~\ceq{coulpots}, it is
 clear that. if
Eq.~\ceq{equalgaia} is
satisfied, the effective Coulomb interactions in ${\cal I}_{4-plat}$
vanish identically, so that the action ${\cal S}_{4-plats}$
coincides once again with
the pure self-dual part ${\cal F}_{4-plat}$.
 Thus, to have  self-duality we only need
that the effects of the
steric interactions are negligible, such as for instance in the case of
very low monomer concentration or at the theta-point.

The self-duality condition
\ceq{equalgaia}
may be physically interpreted as the requirement that the
trajectories $\Gamma_{a,I_a}$ must be homogeneous, in the sense that
they should have equal persistence lengths. In this case, 
the
attractive and repulsive forces described by the term ${\cal
  I}_{4-plat}$ counterbalance themselves in a sort of equilibrium
which establishes the self-dual regime.
This does not mean however that attactions and repulsions between monomers
disappear,
 because Coulomb-like forces still hide in the self-dual part
of the action ${\cal F}_{4-plat}$.

We would like now to exploit further the analogy of the polymer problem
with anyons field theories and search for possible non-trivial
classical solutions
which minimize the polymer free energy ${\cal S}_{4-plat}$.
Of course, one should keep in  mind that,
even in the case of the simplest model of anyons, vortex
solutions outside the self-dual point have been found only by means of
numerical methods. Also at the self-dual point, there is no clear and
detailed understanding of the dynamics of C-S vortices \cite{Dunne}.
Here the situation is further complicated by the presence of replica
fields and multiple families of anyons which are mixed together
due to the constraints (\ref{constrone})-(\ref{constrtwo}).
In view of these limitations, it seems reasonable to restrict
ourselves to the self-duality regime (\ref{equalgaia}) and to static
solutions, which satisfy the conditions
$\partial_t\psi_{a,Ia}^{\omega}=0$ for every value of $a=1,2$,
$I=1,2$ and $\omega=1,\ldots,n_r$.
As a consequence, from now on we will consider the pure self-dual action:
\begin{eqnarray}
&&{\cal S}_{4-plat}=\nonumber\\
&&{\textstyle \frac{(T_2-T_1)}{4g}}\int d\bx\left[
\left|D_+(-\lambda,\bB){\vec \Psi}_{1,1}\right|^2
+
\left|D_+(\lambda,\bB){\vec \Psi}_{1,2}\right|^2+\right.\nonumber\\
&&\left.
\left|D_-(-\frac{\kappa}{2\pi},\bC){\vec \Psi}_{2,1}\right|^2
+
\left|D_-(\frac{\kappa}{2\pi},\bC){\vec \Psi}_{2,2}\right|^2\right]
\label{pureselfdual}
\end{eqnarray}
Of course, static solutions
do not fit very well  with the physical boundary conditions which 
should be imposed to the classical fields at the points of maxima and
minima $T_1$ and $T_2$.
For this reason, we are supposing here implicitly
that polymers are very long and that
the interval $T_2-T_1$ is large. In this way
the relevant contributions to the free energy ${\cal S}_{4-plat}$ come
from instants $t$ which are far enough from the extrema located at
$t=T_1$ and $t=T_2$. At these intermediate points the
trajectories 
fluctuate in such a way that, at each
fixed instant $t$,
they visit more or less with the same frequency the
same locations of the two dimensional space spanned by the vectors
$\br_{a,I_a}(t)$.

From the action (\ref{pureselfdual})
 one finds the
following Euler-Lagrange equations of motion:
\begin{eqnarray}
&&D_+(-\lambda,\bB)\psi_{1,1}^\omega=D_+(\lambda,\bB)\psi_{1,2}^\omega=0
\label{eqoneeqfour}\\
&&D_-\left(-\frac{\kappa}{2\pi},\bC\right)\psi_{2,1}^\omega=D_-\left(
{\displaystyle
\frac{\kappa }{2\pi}},\bC\right)\psi_{2.2}^\omega=0\label{eqfiveeqeight}
\end{eqnarray}
To these equations one should also add the constraints
(\ref{constrone})--(\ref{constrtwo}) which determine the transverse
components of the fields $\bB,\bC$ and the Coulomb gauge fixing
(\ref{coulgauge}). In total, after separating the real and imaginary parts
in Eqs.~(\ref{eqoneeqfour})--(\ref{eqfiveeqeight}), we obtain
  $12\times n_r$
  equations, which completely specify the classical field
  configurations.
It seems natural to try replica symmetric solutions of the kind:
\beq{
\psi_{a,I_a}^\omega=\psi_{a,I_a}\qquad\psi_{a,I_a}^{\omega*}=\psi_{a,I_a}^*
}{replsymm}
where $a=1,2$, $I_a=1,2$ and
$\omega =1,\ldots,n_r$.
Moreover, we switch to polar coordinates in order to express the
complex fields:
\beq{\psi_{a,I_a}=e^{\imath \phi_{a,I_a}}\rho^{1/2}_{a,I_a}}{polcoord}
The consistency of
Eqs.~(\ref{eqoneeqfour})--(\ref{eqfiveeqeight}) requires that:
\beq{
\phi_{a,1}=-\phi_{a,2}\qquad\qquad\rho_{a,1}=
\frac{c_a}{\rho_{a,2}}}{conscond}
Here $c_1$ and $c_2$ are arbitrary constants different from zero.
Also the densities $\rho_{a,I_a}$ must not vanish.
There are other ways to make 
Eqs.~(\ref{eqoneeqfour})--(\ref{eqfiveeqeight}) consistent, but they
lead either to unphysical or to trivial solutions.
After imposing the consistency conditions (\ref{conscond})
in Eqs.~(\ref{eqoneeqfour})--(\ref{eqfiveeqeight}), there remain
still four independent relations, which can be used to fix the
components of the vector fields $\bB,\bC$:
\begin{eqnarray}
\lambda
B_i&=&\partial_i\phi_{1,1}+\frac12\epsilon_{ij}\partial^j\ln\rho_{1,1}
\label{compB} \\
\frac\kappa{2\pi}
C_i&=&\partial_i\phi_{2,1}-\frac12\epsilon_{ij}\partial^j\ln\rho_{2,1}
\label{compC} 
\end{eqnarray}
It is easy to prove that the Coulomb gauge requirement
(\ref{coulgauge}) is satisfied only if the phases
$\phi_{1,1}$ and
$\phi_{2,1}$ are constant.
Finally, inserting Eqs.~(\ref{compB})--(\ref{compC}) in the
constraints (\ref{constrone})--(\ref{constrtwo}), it is possible to
determine the remaining degrees of freedom $\rho_{1,1}$ and
$\rho_{2,1}$:
\begin{eqnarray}
  \Delta\ln\rho_{1,1}&=&4 
\lambda
n_r\left(\frac{c_2}{\rho_{2,1}}-\rho_{2,1}\right)\label{eqnrhoone}\\[.5cm]
  \Delta\ln\rho_{2,1}&=& 4
\lambda
n_r\left(
\rho_{1,1}-\frac{c_1}{\rho_{1,1}}\right)\label{eqnrhotwo}
\end{eqnarray}
A further study of the above two equations requires numerical methods.

Concluding, we have succeded in establishing a connection between
topologically linked polymers and anyon field theories.
We have limited ourselves to $4-$plats, but it is possible to
generalize our results
to the case of any $2s-$plat. In our approach, links which
belong to different $2s-$plat configurations are distinguished. This allows
a better classification of the topological states of polymers
than the bare Gauss linking number.
In view of possible applications to knot theory,
it would be nice to extend also to non-abelian C-S
field theories our techniques based on
the Coulomb gauge and on the sectioning procedure of loops described
above. 

The analogy between polymers and anyons has been exploited to investigate
the statistical mechanics of $4-$plats, a particular class of links,
which is relevant in biological applications. It turns out that,
as  anyon particles in multi-layered systems, also
$4-$plats have a self-dual point. Within the hypothesis of
Eq.~(\ref{equalmaxmin}), which demands that the points of minima
$\tau_{a,2I_a+1}$ are aligned at the same height $t=T_1$, while the
points of maxima are aligned at the height $t=T_2$,
 self-duality
occurs when all open curves $\Gamma_{a,I_a}$ have the same
length. Particular polymer configurations of this kind may exist in nature, 
for instance after
the process of DNA replication.
However, we have seen that the requirements for the presence of
 a self-dual point
are much more general: We only need
that the effects of the
steric interactions are negligible, such as for instance in the case of
very low monomer concentration or at the theta-point.

The existence of the self-dual point \ceq{equalgaia}
shows that the physics of
interacting polymers is much richer than previously
expected. For instance, the classical equations of
motion admit non-trivial solutions, which are the analogs of
vortex-like solutions in anyon models. By analytical methods 
it is just
possible to investigate static field configurations. These
suffer the limitations mentioned after
Eq.~(\ref{pureselfdual}) and cannot be defined if the requirements of
Eqs.~(\ref{equalmaxmin}) are not fulfilled.
In fact, 
if maxima and minima are localized at different heights on
the $t-$axis, the
action ${\cal S}_{4-plat}$ becomes explicitly dependent
on time
due to the presence of the
Heaviside $\theta-$functions in the fields $\bB$ and $\bC$.
This time dependence is very mild. Its only effect is that the
boundaries of integration over
 time in the various terms composing the action ${\cal
  S}_{4-plat}$  do not coincide. Yet, this is sufficient to spoil the
 possibility of having purely static field configurations.
For these reasons, any conclusion on the meaning of the static solutions
in the case of polymers
should be taken with some care. Having in mind these caveats, it is
interesting to note that, in the solutions which we have studied here,
it seems to prevail a repulsive force between couples of trajectories
$\Gamma_{a,I_a}$ belonging to the same polymer. This is
what  Eq.~(\ref{conscond}) suggests, because 
the densities $\rho_{a,1}$ are constrained to be inversely proportional 
to their
counterparts $\rho_{a,2}$ at each point
$\bx$.

It is worthing to point out that, despite their close resemblances,
the mechanisms used to establish
 the self-dual regime 
in our polymer model and in usual anyon field theories
are different.
In anyon field theories, in fact, the existence of the self-dual point
requires that the effective Coulomb interactions, which arise after
 performing
the  Bogomol'nyi transformations,
are canceled against true Coulomb
 interactions of charged particles \cite{Ezawa}.
In the polymer model, instead, the
 effective Coulomb interactions,
represented by the 
 term ${\cal I}_{4-plat}$, 
 counterbalance themselves and the
 presence of
other interactions is not needed. This self-balancing transition to
 self-duality may occur also in other physical problems where
 multi-layered systems of non-charged particles are relevant.
Another application of our approach are polymer brushes. In that case,
 C-S field theories in the Coulomb gauge are able to take
 into account the effects related to the winding
of neighboring
 polymer trajectories.

\begin{acknowledgments}
Part of this work has been carried out during a
visit
 to the Max Planck Institute for Polymer Research. This visit
 has been funded by a grant from 
the German Academic Exchange Service (DAAD), which is gratefully acknowledged.
The author wishes also to
thank the Max Planck Institute for Polymer Research
for the nice hospitality.
He is also indebted to V. A. Rostiashvili and T. Vilgis, which 
participated
in the early stages of this work,
for their invaluable
help and their constant encouragement. 
\end{acknowledgments}

\end{document}